# Stability and Queueing Analysis of IEEE 802.11 Distributed Coordination Function

Dongjie Yin, Pui King Wong, and Tony T. Lee

Abstract — A widely adopted two-dimensional Markov chain model of the IEEE 802.11 DCF was introduced by Bianchi to characterize the backoff behavior of a single node under a saturated traffic condition. Using this approach, we propose a queuing model for the 802.11 DCF under a non-saturated traffic environment. The input buffer of each node is modeled as a Geo/G/1 queue, and the packet service time distribution is derived from Markov state space of 802.11 DCF with the underlying scheduling algorithm. The DCF defines two access mechanisms, namely the Basic access mechanism and the request-to-send/clear-to-send (RTS/CTS) access mechanism. Based on our model, performance analyses of both schemes are studied with probabilistic exponential backoff scheduling. We obtain the characteristic equation of network throughput and expressions of packet queueing delay. Specifically, we obtain the stable throughput and bounded delay regions with respect to the retransmission factor according to the basic queueing analysis. For both access schemes, the bounded delay region is a subset of the stable throughput region. Our results show that the RTS/CTS access mechanism is more stable and performs better than the Basic access mechanism. The analysis in this paper is verified by simulation results.

*Keywords: performance analysis, queuing analysis, IEEE 802.11 DCF protocols, scheduling.*

## I. INTRODUCTION AND OVERVIEW

The standardized Media Access Control (MAC) protocol plays an important role in wireless local area networks (WLANs). The IEEE 802.11 protocol with distributed coordination function (DCF) is the most popular standard that includes specifications for both MAC and physical layers. The core of the 802.11 MAC protocol is the Carrier-Sense Multiple-Access protocol with Collision Avoidance (CSMA/CA). A node with a fresh packet monitors the channel activity first. If the channel is sensed idle for a period of time, called *Distributed Inter-Frame Space* (DIFS), the node transmits the packet. Otherwise, the node persists monitoring the channel activity until sensing an idle channel for a DIFS period, and then generates a random backoff interval before transmission.

The DCF defines two access mechanisms, namely the Basic access mechanism and the request-to-send/clear-to-send (RTS/CTS) access mechanism. The Basic access mechanism is simply a two-way handshaking scheme, in which the destination node acknowledges the successful receipt by sending an ACK frame. On the other hand, the RTS/CTS access mechanism uses a four-way handshaking method to eliminate the hidden node problem. Before sending the packet payload, the source node broadcasts a request-to-send (RTS) short message to request channel resources. If no collision occurs, the destination node sends a clear-to-send (CTS) short message to indicate that the channel is reserved. The source node starts to transmit the packet payload after receiving the CTS. The source node also receives an ACK frame if the packet payload is successfully transmitted.

In addition, the DCF employs the window-based exponential backoff scheduling scheme. The backoff time is uniformly chosen from the range of (0, CW-1), in which the window size CW, called the *Contention Window*, is determined by the current backoff phase. The backoff time counter is decremented if the channel is sensed idle, "frozen" if the channel is busy, and reactivated when the channel is sensed idle again for a DIFS period. The packet is ready to be sent when the counter reaches 0.

The model of the 802.11 DCF protocol proposed by Bianchi in [1] was a groundbreaking work in the throughput analysis of MAC protocols in conjunction with the backoff algorithm. The analysis assumed that nodes always have packets to transmit. This saturation assumption substantially simplified the analysis by ignoring packet arrival processes, but failed to comprehend the complete queuing behaviors of a single node, such as stability and delay, issues concerning regular network operation under non-saturated condition.

Abundant published works on the performance of 802.11 protocols under different scenarios (see [2–4]) were stimulated by Bianchi's two-dimensional Markov chain model. In [2], the authors used the same model to derive the network throughput with retry limit. The saturation throughput for a fading channel by considering the probability that packets collapsed due to frame error is studied in [3], which resulted in smaller network throughput as expected because of the larger probability of collapsed transmissions. Moreover, the authors in [4] applied Bianchi's two-dimensional Markov chain model to analyze the network throughput of an alteration of the 802.11 protocol. Again, these works are all limited to throughput analysis due to the saturated traffic assumption.

Several approaches different from the two-dimensional Markov chain were proposed in [5–8] to analyze the access delay of 802.11 networks. The service time distribution was estimated by simulation in [5]; while the access delay for the 802.11 with binary exponential backoff (BEB) was studied by using a bottom-up approach in [6]. Furthermore, the access delay of the 802.11 DCF protocol has also been studied in [7] and [8] by considering the generating function of packet service time distribution. Nevertheless, these approaches were all based on the saturated traffic assumption.

Recently, some extensions of Bianchi's model to a non-saturated environment were reported in [9–14]. An extra idle state of the Markov chain is introduced in [9–12] to represent the empty queue after a successful packet transmission. A system without input buffers was studied in [9], the buffer-less assumption can help to avoid the complicated queuing analysis, but the result is overly simplified and not useful in practice. An analytical model with small buffers for 802.11 DCF in the presence of unsaturated heterogeneous conditions was proposed in [10] and [11], in which the characteristic equation of network throughput was found to be the same as that under the saturated condition. Liaw et al. has derived a good approximation for analyzing an unsaturated system in [12], in which the probability that the buffer is non-empty after a successful transmission is assumed to be a constant and independent of the access delay of the transmitted packet. Unfortunately, these assumptions are too restrictive for analyzing the queuing delay of an unsaturated system in standard queuing theory. The existing queueing analyses of the IEEE 802.11 DCF either failed to completely model the standard specifications of 802.11 DCF protocol, or depended on idealized hypotheses and approximately estimated parameters [13–16]. As a consequence, these analytically results drastically deviated from the simulations, and generally neglected the stability issue of network throughput.

In light of the above concerns, we propose a generic queuing model of the 802.11 DCF, not only to derive the network throughput and packet delay, but also to tackle the stability issues. As the 802.11 protocol is being continuously updated to meet the requirements of different types of emerging new services, a model of the MAC protocol should be flexible enough to cope with the changes of QoS requirements. This queueing model is an outgrowth of previous analyses of CSMA protocols reported in [17] and [18], in which each input buffer is modeled as a Geo/G/1 queue with a Bernoulli arrival process. The service time distribution is derived by a Markov chain describing the state transitions of head-of-line (HOL) packets. We use the probabilistic backoff scheme, in which the retransmission of backlogged packets is determined by the probabilistic retransmission factor, to model the contention window used by Bianchi et al. The probabilistic backoff scheme has been widely applied to the analysis of MAC protocols for its mathematical simplicity [19–21]. The results provided in Appendix II show that the network throughput derived from our model coincides with that of window-based backoff scheme reported in [1], [10] and [11].

Our main contribution is to specify the stability conditions of the 802.11 networks in terms of throughput and delay, which were largely ignored in other related works with window-based backoff scheme. Specifically, we obtain the stable throughput region as well as the bounded delay region for the Basic and RTS/CTS access mechanisms of IEEE 802.11 DCF with the exponential backoff scheme. We show that the stable condition of throughput can be determined by the first moment of service time, or offered load, while the bounded delay condition requires a bounded second moment of service time according to the Pollaczek-Khinchin (P-K) formula of Geo/G/1 queue [22]. In addition, the simulation results exhibited that the queueing delay derived from our Markov state space analysis of DCF is much more accurate in comparison with that in [13–16].

In general, for a system operating inside the stable throughput region but not within the bounded delay region, a stable network throughput can still be achieved, but the queuing delay may be unacceptably high due to the large variance of packet service time. We also prove that the stable throughput region still exists even for an infinite population, which agrees with that previously reported in [23] and [24]. Moreover, with the RTS/CTS access mechanism, the network performance is less dependent on the aggregate input rate and retransmission factor, and offers a higher achievable throughput than that of the Basic access mechanism.

The rest of this paper is organized as follows. In section II, a Markov chain of HOL packets under the 802.11 protocol is proposed to carry out the derivation of network throughput and packet service time distribution for the queuing model of the input buffer. The regions of the stable throughput and bounded queuing delay are described in sections III and IV, respectively, for both access mechanisms with the exponential backoff scheme. A conclusion is given in section V.

## II. QUEUING MODEL OF HOL PACKET FOR THE 802.11 DCF

The theme of this paper is to provide a complete stability and queueing analysis of 802.11 DCF networks. We consider the network as a multi-queue system with a single server. The services of HOL packets of input buffers depend on the behavior of the channel. In this section, we first describe the channel status in the context of an alternating renewal process, and then model the service time distribution of HOL packets as a Markov chain with state transitions defined by the 802.11 protocol.

### A. *Alternating Renewal Process of Channel*

The network under consideration is governed by the 802.11 DCF. We assume that the time axis consists of mini-slots with slot size $a$ and that the network is synchronized. Packets can be sent only at the beginning of a mini-slot. The channel status shown in Fig. 1 is an alternating sequence of busy and idle periods. The length of an idle period is a random variable with geometric distribution as in [25]. Since the channel is idle in the DIFS period following any transmission, a transmission period comprises a packet transmission and the subsequent DIFS. The busy period is a series of consecutive transmission periods, in which either a successful transmission or a collision has occurred.

Let $G$ be the aggregate attempts generated by all fresh and re-scheduled HOL packets in one time slot. According to the theorem on the superposition of point processes proved in [26], the aggregate attempts form a Poisson process for large population. This Poisson assumption has been firstly adopted in [22] and [25], the seminal papers on the analysis of MAC protocols. Let $P_t$ be the probability that there is at least one attempt of transmission in a mini-slot. The following events may occur in a mini-slot when the channel is ready for accepting requests:

1) The channel remains in the idle period if no one makes a request with the following probability:

$$1 - P_t = e^{-aG}. \tag{1}$$

2) A transmission period is initiated by at least one request with probability $P_t$. The transmission period can be either successful with probability $P_S$ that only one request is generated:

$$P_s = \frac{aGe^{-aG}}{P_t} = \frac{aGe^{-aG}}{1 - e^{-aG}}, \tag{2}$$

or collided with probability $1 - P_s$.

Thus, we have the following probabilities in respect to the channel status in a mini-slot when the channel is ready:

i. The channel is idle in the mini-slot with probability $\pi_{Idle} = 1 - P_t$.

ii. The mini-slot leads a successful transmission period with probability $\pi_{Suc} = P_t P_s$.

iii. The mini-slot leads a collision period with probability $\pi_{Col} = P_t(1 - P_s)$.

From the alternating sequence of busy and idle periods shown in Fig. 1, the probability $\alpha$ that the channel is available for accepting request in a mini-slot is given as follows:

$$\alpha = \frac{a\pi_{Idle} + a\pi_{Suc} + a\pi_{Col}}{a\pi_{Idle} + T_S\pi_{Suc} + T_C\pi_{Col}} = \frac{a}{a(1-P_t) + T_S P_t P_s + T_C P_t (1-P_s)}, \qquad (3)$$

where $T_S$ and $T_C$ are respective lengths of a successful transmission period and a collision period.

Based on the above description of channel status, the following section describes a detailed queuing model of an individual input buffer in which the service time distribution is derived in conjunction with the exponential backoff scheme.

B.  *Queuing Model of Input Buffer*

Consider an 802.11 DCF network with *n* nodes, the input buffer of each node is modeled as a Geo/G/1 queue with Bernoulli arrival process. The *K*-exponential backoff algorithm is employed for contention resolutions as described in [17] and [18]. A fresh HOL packet is initially in phase 0, which will be increased by 1 each time the packet encountered a collision. A backlogged HOL packet in phase *i*, for $i = 1, \ldots, K$, has retransmission probability $q^i$, where *q* is a retransmission factor chosen from the interval (0, 1) and *K* is the cut-off phase.

The Markov chain shown in Fig. 2 describes the state transitions of an HOL packet. Besides the initialization state Act (action state) and the termination state Suc (successful transmission state), the HOL packet in phase *i* can be in one of the three fundamental states: sensing ($S_i$), collision ($Col_i$), or waiting ($W_i$), where $i = 0, \ldots, K$. A fresh HOL packet in phase 0 will be sent only when the channel is being sensed idle for a DIFS period of length $T_D$. The state of the packet in the first ($T_D - a$) mini-slot time of this initial sensing DIFS period is represented by the action state Act. The last mini-slot time of DIFS is sensing state $S_0$.

A fresh HOL packet in sensing state $S_0$ will be immediately transmitted if the channel is available. It moves to the state Suc with a probability of success *p*, or to the collision state $Col_0$ otherwise. The collided packet instantly moves to the next sensing state $S_1$ and repeats the access process. If the channel is sensed busy during a DIFS period, the HOL packet becomes a backlogged packet and moves to the waiting state $W_0$. The packet must wait until the channel becomes idle for a duration of $(T_D - a)$ before it can be moved to the next sensing state $S_1$. A backlogged packet in phase $i$ repeats the same access procedure, except that the packet will be transmitted with a probability $q^i$ in the sensing state $S_i$ when the channel is available.

The transition probability $\alpha_0$ of the Markov chain shown in Fig. 2 is the probability that the channel remains idle for a duration of $(T_D - a)$. Since idle mini-slots are identical and independent, it follows from (3) that

$$\alpha_0 = \alpha^{(T_D-a)/a}. \tag{4}$$

The transmission of a tagged HOL packet can be successful only if all the other nodes do not send packets in the same mini-slot. Hence, the probability of successful transmission $p$ is given as follows:

$$p = e^{-aG}. \tag{5}$$

Let $b_{s_i}, b_{col_i}, b_{w_i}, b_{act}$, and $b_{suc}$ be the respective limiting probabilities of states $S_i$, $Col_i$, $W_i$, Act, and Suc of the Markov chain. From the Markov chain shown in Fig. 2, we obtain the following set of state equations:

$$\begin{aligned}
&b_{suc} = \alpha p b_{s_0} + \alpha p \sum_{i=1}^{K} b_{s_i} q^i, \; b_{act} = b_{suc}, \; b_{w_0} = (1-\alpha_0) b_{act} + (1-\alpha) b_{s_0}, \\
&b_{col_0} = \alpha(1-p) b_{s_0}, \; b_{s_0} = \alpha_0 b_{act}, \; b_{s_1} = \alpha(1-q) b_{s_1} + b_{w_0} + b_{col_0} + b_{w_1}, \\
&\begin{cases} b_{col_i} = \alpha q^i (1-p) b_{s_i} \\ b_{w_i} = (1-\alpha) b_{s_i} \end{cases} \text{for } i = 1, 2, \cdots, K \\
&b_{s_i} = \alpha(1-q^i) b_{s_i} + b_{col_{i-1}} + b_{w_i} \quad \text{for } i = 2, 3, \cdots, K-1 \\
&b_{s_K} = \alpha(1-q^K) b_{s_K} + b_{col_{K-1}} + b_{col_K} + b_{w_K}.
\end{aligned} \tag{6}$$

It is easy to prove from (6) that if $p + q > 1$, then all states of the Markov chain are positive recurrent and aperiodic [27]. The HOL packet is in a waiting state if the channel is busy in transmission, either successful or collided. Since the packet must stay in the waiting state until the channel becomes idle for a period of $(T_D - a)$, the duration of waiting states $W_i$, denoted as $T_W$ for all phase $i$, is given by

$$T_W = pT_S + (1-p)T_C - a. \tag{7}$$

The sojourn times of states Act, Suc, $S_i$, Col$_i$, and $W_i$ are $t_{Act} = T_D - a$, $t_{Suc} = T_S - T_D$, $t_{S_i} = a$, $t_{Col_i} = T_C - a$, and $t_{W_i} = T_W$, respectively. The time-average probabilities of those states can be easily determined from the state equations given in (6) as follows:

$$\tilde{b}_{act} = (T_D - a)\alpha p D^{-1}; \tilde{b}_{suc} = (T_S - T_D)\alpha p D^{-1}$$

$$\tilde{b}_{S_i} = \begin{cases} a\alpha_0 \alpha p D^{-1} & \text{for } i=0 \\ ap(1-p\alpha\alpha_0)(1-p)^{i-1}q^{-i}D^{-1} & \text{for } i=1,...,K-1 \\ a(1-p\alpha\alpha_0)(1-p)^{K-1}q^{-K}D^{-1} & \text{for } i=K \end{cases}$$

$$\tilde{b}_{w_i} = \begin{cases} T_W \alpha p (1-\alpha\alpha_0) D^{-1} & \text{for } i=0 \\ T_W p(1-\alpha)(1-p\alpha\alpha_0)(1-p)^{i-1}q^{-i}D^{-1} & \text{for } i=1,...,K-1, \\ T_W (1-\alpha)(1-p\alpha\alpha_0)(1-p)^{K-1}q^{-K}D^{-1} & \text{for } i=K \end{cases} \tag{8}$$

$$\tilde{b}_{col_i} = \begin{cases} \alpha^2 \alpha_0 p(T_C - a)(1-p)D^{-1} & \text{for } i=0 \\ \alpha p(T_C - a)(1-p\alpha\alpha_0)(1-p)^{i}D^{-1} & \text{for } i=1,...,K-1 \\ \alpha(T_C - a)(1-p\alpha\alpha_0)(1-p)^{K}D^{-1} & \text{for } i=K \end{cases}$$

where

$$D = (T_W - \alpha_0 \alpha T_W + T_S + a\alpha_0 - T_C)\alpha p + (T_C - a)\alpha$$
$$+ \frac{(1-p\alpha_0\alpha)(a+T_W-\alpha T_W)}{q(p+q-1)}\left[pq + (p+q-1-pq)\left(\frac{1-p}{q}\right)^{K-1}\right]. \tag{9}$$

The input rate $\lambda$ for each node is defined as the number of arrivals per duration of $t_{Suc}$, where $t_{Suc}$ is the sojourn time of the Suc state shown in Fig. 2. Since the Markov chain under consideration is positive recurrent, the mean service time of an HOL packet scaled by $t_{Suc}$ is the mean return time $1/\tilde{b}_{suc}$ of the state Suc. It follows from (8) that the offered load $\rho$ of each input queue is given as follows:

$$\rho = \lambda/\tilde{b}_{suc} = \lambda D/(t_{Suc}\alpha p), \tag{10}$$

which is also interpreted as the probability that an input buffer is non-empty. From the queuing model of the input buffer, we derive the characteristic equation of network throughput for the 802.11 DCF in the following theorem.

**Theorem 1.** For the 802.11 DCF with the general exponential backoff algorithm, the throughput in equilibrium is given by

$$\hat{\lambda}_{out} = \frac{-p\ln(p)E[P]}{(1-P_t)a + P_t P_S T_S + P_t(1-P_S)T_C} = \frac{aGe^{-aG}E[P]}{e^{-aG}a + aGe^{-aG}T_S + (1-e^{-aG} - aGe^{-aG})T_C}. \quad (11)$$

**Proof:** In the 802.11 protocol, a node is ready to send HOL packets only if the channel is idle in the mini-slot of a sensing state. The probability of successful transmission from a desired node is the conditional probability that none of other nodes access the channel, given that all nodes sense the channel idle,

$$p = \Pr\{\text{none of other } n-1 \text{ nodes access the channel} \mid \text{channel is sensed idle}\}$$
$$= \frac{\Pr\{\text{none of other } n-1 \text{ nodes access the channel}\}}{\Pr\{\text{channel is sensed idle}\}}. \quad (12)$$

The event that no one is attempting to access the channel occurs if all the other $n-1$ nodes are in one of the three cases: i. It is empty; ii. It is in the action state Act; iii. It is in sensing states but not attempting to send packet.

Thus, we have

$$\Pr\{\text{none of other } n-1 \text{ nodes access the channel}\}$$
$$= \left[\Pr\{\text{empty}\} + \Pr\{\text{in action state}\} + \Pr\{\text{in sensing state but not attempting to send packet}\}\right]^{n-1} \quad (13)$$
$$= \left[(1-\rho) + \rho \cdot \tilde{b}_{act} + \rho \sum_{i=1}^{K} \tilde{b}_{s_i}(1-q^i)\right]^{n-1} \stackrel{\text{for large n}}{=} \exp\left\{-n\rho\left[1 - \tilde{b}_{act} - \sum_{i=1}^{K} \tilde{b}_{s_i}(1-q^i)\right]\right\}.$$

Similarly, the probability that all nodes sense an idle channel is given by:

$$\Pr\{\text{channel is sensed idle}\} = \left[\Pr\{\text{empty}\} + \Pr\{\text{in action state}\} + \Pr\{\text{in sensing state}\}\right]^{n-1}$$
$$= \left[(1-\rho) + \rho \cdot \tilde{b}_{act} + \rho \sum_{i=0}^{K} \tilde{b}_{s_i}\right]^{n-1} \stackrel{\text{for large n}}{=} \exp\left\{-n\rho\left[1 - \tilde{b}_{act} - \sum_{i=0}^{K} \tilde{b}_{s_i}\right]\right\}. \quad (14)$$

Then the probability of successful transmission $p$ defined by (12) can be expressed as:

$$p = \exp\{-\lambda na/(t_{suc} \alpha p)\}. \quad (15)$$

Substituting (3) into (15), we have

$$\hat{\lambda} = \lambda n = \frac{-t_{suc} p \ln p}{a(1-P_t) + T_S P_t P_S + T_C P_t (1-P_S)}, \quad (16)$$

where $\hat{\lambda} = n\lambda$ is the aggregate input rate of the entire network. Since only a fraction of $t_{Suc}$ is the packet payload $E[P]$ as presented in [1], the network throughput is given by

$$\hat{\lambda}_{out} = \lambda n \cdot E[P]/t_{Suc}. \quad (17)$$

∎

The network throughput in equilibrium given in (11) is derived without the saturation assumption, yet it is consistent with that of Bianchi's results [1] for saturated networks. Thus, the characteristics of network throughput remain the same in spite of input traffic conditions, which agrees with the studies reported in [10] and [11]. Moreover, the fact that the above derivation is invariant with respect to the retransmission factor $q$ of the backoff scheme suggests that the throughput analysis is too primitive to describe the stability of the network. To fully comprehend the performances of 802.11 DCF, we resort to the Geo/G/1 queuing model of each input buffer.

The key to the model of the input buffer is the moment generating function of packet service time distribution. Let random variables $Act^*, Suc^*, S_i^*, Col_i^*$, and $W_i^*$ be the service completion time of an HOL packet, starting from the state Act, Suc, $S_i$, $Col_i$, and $W_i$, respectively, until it is successfully transmitted. We assume without loss of generality that $t_S = t_{Suc}/a$, $t_A = t_{Act}/a$, $t_C = t_{Col_i}/a$, and $t_W = t_{W_i}/a$ are integers. From the Markov chain shown in Fig. 2, the generating functions $Act(z)$, $Suc(z)$, $S_i(z)$, $Col_i(z)$, and $W_i(z)$ of these service completion times satisfy the following set of equations:

$$Act(z) = E\left[z^{Act^*}\right] = \alpha_0 z^{t_A} S_0(z) + (1-\alpha_0) z^{t_A} W_0(z)$$

$$Suc(z) = E\left[z^{Suc^*}\right] = z^{t_S}$$

$$S_i(z) = E\left[z^{S_i^*}\right] = \begin{cases} \alpha p z Suc(z) + \alpha(1-p) z Col_0(z) + (1-\alpha) z W_0(z) & i=0 \\ \alpha q^i p z Suc(z) + \alpha q^i (1-p) z Col_i(z) \\ + (1-\alpha) z W_i(z) + \alpha(1-q^i) z S_i(z) \end{cases} \quad i=1,\ldots,K.$$

$$Col_i(z) = E\left[z^{Col_i^*}\right] = \begin{cases} z^{t_C} S_{i+1}(z) & i = 0, ..., K-1 \\ z^{t_C} S_K(z) & i = K \end{cases}$$

$$W_i(z) = E\left[z^{W_i^*}\right] = \begin{cases} z^{t_W} S_1(z) & i = 0 \\ z^{t_W} S_i(z) & i = 1, ..., K \end{cases} \quad (18)$$

Let $X$ be the service time of an HOL packet. The first and second moments of service time, $E[X]$ and $E[X^2]$, can be derived from the generating function $Act(z)$ and are given in Appendix I, because the service of each HOL packet always starts from the action state Act. These moments of service time are vital to the performance analysis covered in section IV.

III. STABLE THROUGHPUT REGION FOR THE 802.11 DCF

In this section, we study the region of the retransmission factor $q$ in which the 802.11 DCF with the exponential backoff scheme has stable network throughput, i.e. the channel's output rate equals the aggregate input rate. The Basic access mechanism and RTS/CTS access mechanism are considered. We first specify the stable condition subject to the throughput characteristics given in (11), and then the region of retransmission factor $q$ is determined from the particular retransmission rule of exponential backoff.

A. *Stable Throughput Condition*

The Basic and RTS/CTS access mechanisms of DCF can be distinguished by the values of parameters $T_S$ and $T_C$ listed in Table. 1 as defined in [1]. Based on these parameters, the network throughput specified by (11) is plotted in Fig. 3 for both access mechanisms, in which $\hat{\lambda}_{max}$ denotes the maximum throughput. Both curves, as depicted in Fig. 3, first increase and then decrease versus the attempt rate $G$. Hence, the throughput equation (11) has two roots, denoted as $G_S(\hat{\lambda}_{out})$ and $G_L(\hat{\lambda}_{out})$, respectively, for any throughput $\hat{\lambda}_{out} < \hat{\lambda}_{max}$.

The optimal throughput reveals the intrinsic tradeoff between the attempt rate $G$ and probability of success $p = e^{-aG}$. To achieve a stable throughput $\hat{\lambda}_{out} = \hat{\lambda} E[P]/t_{Suc} < \hat{\lambda}_{max}$, the attempt rate $G$ cannot be too small or too large. As shown in Fig. 3, the network capacity is higher than the required throughput only if the attempt rate $G$ lies between the two roots $G_S(\hat{\lambda}_{out})$ and $G_L(\hat{\lambda}_{out})$. Hence, a necessary condition of stable throughput of the entire system is defined as follows:

**Stable Throughput Condition (STC):** *For any aggregate input rate $\hat{\lambda}$ with the stable throughput $\hat{\lambda}_{out} = \hat{\lambda} E[P]/t_{Suc} < \hat{\lambda}_{max}$, the attempt rate $G$ should satisfy*

$$G_S(\hat{\lambda}_{out}) \leq G \leq G_L(\hat{\lambda}_{out}). \tag{19}$$

In the following discussions, we simply use $G_S$ and $G_L$ to denote the two roots of (11) with the understanding that they are functions of the network throughput $\hat{\lambda}_{out}$.

Comparing the two throughput curves plotted in Fig. 3, similar to those depicted in [1], the maximum throughput of the Basic access mechanism is very close to that of the RTS/CTS access mechanism. However, the throughput of the Basic scheme is more sensitive to the change of attempt rate $G$. In fact, as Fig. 3 shows, a small variation in the value of $G$ leads to a larger fluctuation in the throughput for the Basic access mechanism. For a given input rate $\hat{\lambda} = 0.3$, the network with the Basic access mechanism has a stable throughput $\hat{\lambda}_{out} = \hat{\lambda} E[P]/t_{Suc} \approx 0.277$ when the attempt rate $G$ is within the range [≈ 0, ≈ 0.042]; while that of the RTS/CTS access mechanism is approximately equal to 0.260 when $G$ is in between [≈ 0, ≈ 0.115].

B. *Stable Throughput Region of Exponential Backoff*

In this section, we first establish the relationship between the attempt rate $G$ and the retransmission factor $q$ of exponential backoff, and then determine a stable throughput region of $q$ from the **STC** on the attempt rate $G$.

For an 802.11 DCF network with $n$ nodes, the desired HOL packet can be transmitted only when all the other nodes are inactive in the mini-slot prior to its transmission. Hence, we only focus on the attempt rate $G$ in this mini-slot in this stability analysis. Suppose that there are a total of $n_b = \sum_{i=1}^{K} n_i$ busy nodes in a mini-slot, each contains backlogged HOL packets, in which $n_i$ packets are in the sensing state of phase $i$, for $i = 1,\ldots, K$. We consider the following two cases in the mini-slot before a transmission period:

1) A packet arrives at an empty node in the mini-slot with probability $a\lambda/t_{Suc}$, and there are $(n - n_b)$ empty nodes. Recall that the input rate $\lambda$ is defined as the probability that a packet arrives in duration of $t_{Suc}$, and the newly arrived packet will be sent with probability $\alpha_0 \alpha$, that the channel is idle for a DIFS period;

2) A backlogged HOL packet in phase $i$ will be transmitted with probability $q^i$, for $i = 1,\ldots, K$.

Collectively, the attempt rate $aG$ in this mini-slot can be expressed as follows:

$$aG = (\alpha_0 \alpha a \lambda / t_{Suc}) E[n - n_b] + \sum_{i=1}^{K} q^i E[n_i]. \tag{20}$$

The mean number of empty nodes in the system can be expressed as

$$E[n - n_b] = n(1 - \rho); \tag{21}$$

while the mean number of nodes in phase $i$ is

$$E[n_i] = n\rho \tilde{b}_{s_i} \Big/ \sum_{j=1}^{K} \tilde{b}_{s_j}. \tag{22}$$

Substituting (21) and (22) into (20), we obtain the following attempt rate in the mini-slot:

$$aG = n(1-\rho)\alpha_0 \alpha a \lambda / t_{Suc} + n\rho \sum_{i=1}^{K} \tilde{b}_{s_i} q^i \Big/ \sum_{j=1}^{K} \tilde{b}_{s_j}. \tag{23}$$

Finally, the attempt rate in the mini-slot for exponential backoff ($K = \infty$) can be obtained by substituting (8) and (10) into (23) and is given by

$$aG = (1-\rho)\alpha_0 \alpha a \hat{\lambda} / t_{Suc} + n\rho(p + q - 1)/p. \tag{24}$$

The attempt rate $G$ is actually an implicit function of the retransmission factor $q$ associated with the underlying scheduling algorithm. From equation (24), the retransmission factor $q$ can be formulated in terms of attempt rate $G$ as shown below:

$$q = h(G) = (1-p) + \sqrt{\left(\frac{B(p,n)}{2\hat{\lambda}A(p)}\right)^2 + \frac{\hat{\lambda}a\alpha_0 p(1-\alpha\alpha_0 p)(a+T_W-\alpha T_W)}{A(p)n(T_S-T_D)}} - \frac{B(p,n)}{2\hat{\lambda}A(p)}, \quad (25)$$

where
$$A(p) = T_W(1-\alpha\alpha_0) + T_S + a\alpha_0 - T_C + (T_C - a)p^{-1} \quad (26)$$

and $B(p,n) = (n - \hat{\lambda}A(p))\alpha\alpha_0 a\hat{\lambda}pn^{-1} + \hat{\lambda}\alpha^{-1}(1-\alpha\alpha_0 p)(a+T_W-T_W\alpha) - aGp(T_S-T_D).$ (27)

The function $h(G)$ in (25) monotonically increases with respect to $G$, which implies that the **STC** given in (19) precisely determines the following stable throughput region $R_T$ of the retransmission factor $q$,

$$q \in R_T = [h(G_S), h(G_L)]. \quad (28)$$

The stable throughput regions $R_T$ for the Basic and RTS/CTS access mechanisms are areas shown respectively in Fig. 4a and Fig. 4b with $n = 10$. For aggregate input rate $\hat{\lambda} = 0.3$, the corresponding stable throughput regions of $q$ are [≈ 0.049, ≈ 0.875] and [≈ 0.0478, ≈ 1] for the Basic and RTS/CTS schemes. These figures also demonstrate that the maximum throughput is achieved at $G_S(\hat{\lambda}_{max}) = G_L(\hat{\lambda}_{max})$ when the stable region of retransmission factor $q$ shrinks to a single point.

In the extreme case, as the number of nodes $n$ goes to infinity, the expression (25) becomes

$$q = 1 - p. \quad (29)$$

Then the following non-empty stable throughput region $R_T$ can be obtained from (28) and (29),

$$R_T = [1 - e^{-aG_S}, 1 - e^{-aG_L}], \quad (30)$$

which coincides with Song's results proved in [23] and [24] that the network throughput can be non-zero for exponential backoff even for an infinite number of nodes.

The simulation results shown in Fig. 5, with $n = 10$ and $\hat{\lambda} = 0.3$, confirm our stability analysis that the stable throughput can be achieved if the retransmission factor $q$ is properly chosen from the stable throughput region. For the Basic access mechanism, the network throughput gradually descends when the retransmission factor is chosen outside this region. On the other hand, the stable throughput region of the RTS/CTS access mechanism covers almost the entire range of the (0, 1) interval. The throughput decreases only when the retransmission factor $q$ closely approaches either 0 or 1.

## IV. BOUNDED DELAY REGION FOR IEEE 802.11

The stable throughput is conditioned on the first moment of service time. For the Geo/G/1 model of input buffers, it is not sufficient to guarantee the bounded mean delay of packets. In this section, we deduce some additional constraint on the retransmission factor $q$ from the second moment of service time. This new constraint gives rise to the specification of the bounded delay region for the 802.11 DCF with exponential backoff.

### A.  *Bounded Delay Condition*

Recall that the mean service rate of packets for each node is $\tilde{b}_{suc}$ per duration of $t_{Suc}$ as in (10). A generic expression of the network throughput is given by $\hat{\lambda}_{out} = \min\{n\tilde{b}_{suc}, \hat{\lambda}\} \cdot E[P]/t_{Suc}$. The stable throughput condition **STC** ensures $\hat{\lambda}_{out} = \hat{\lambda} E[P]/t_{Suc}$, which implies $n\tilde{b}_{suc} \geq \hat{\lambda} = n\lambda$, or equivalently, the offered load $\rho \leq 1$. On the other hand, it can be shown from (10) that $\rho$ monotonically increases with respect to the retransmission factor $q$ if the attempt rate is bounded in the range $G_S \leq G \leq G_L$. In particular, the attempt rate $G$ will reach $G_L$ when the offered load $\rho$ = 1. That is, the condition $q \in [h(G_S), h(G_L))$ is equivalent to the stable condition $\rho = \lambda' E[X] < 1$ of Geo/G/1 queue, where $\lambda' = \lambda/t_{Suc}$ is the scaled input rate of an input buffer. Thus, the **STC** only ensures that the mean service time $E[X]$ is bounded. To guarantee bounded mean queuing delay of the packet, the service time distribution should also satisfy the following higher moment condition.

**Bounded Delay Condition (BDC):** *The Pollaczek-Khinchin formula for mean queuing delay E[T]*

*of Geo/G/1 queue* [22]

$$E[T] = E[X] + \frac{\lambda' E[X^2] - \lambda' E[X]}{2(1 - \lambda' E[X])}. \tag{31}$$

*requires bounded second moment of service time* $E[X^2] < \infty$.

The condition **BDC** is more restrictive than the stable throughput condition **STC**. It is obvious that the delay stability implies the throughput stability, but the converse is not necessarily true. Detailed discussions are provided below.

B. *Bounded Delay Region of Exponential Backoff*

The second moment of service time of the exponential backoff scheme is presented in (36) of Appendix I as follows:

$$E[X^2] = C(p,q) \lim_{K \to \infty} \left[ (1-p)/q^2 \right]^{K-1} + D(p,q), \tag{32}$$

where $C(p, q)$ and $D(p, q)$ are two polynomials given in (37) and (38), respectively, of Appendix I. The **BDC** requires the convergence of the term $\lim_{K \to \infty} \left[ (1-p)/q^2 \right]^{K-1}$ in (32), which implies

$$q > \sqrt{1-p} = \sqrt{1-e^{-aG}}. \tag{33}$$

This condition (33) is consistent with that given by Yang and Yum for the bounded delay with binary exponential backoff in [28]. The second moment of service time only provides a lower bound of the retransmission factor $q$. The bounded delay region $R_D$ shares the same upper bounded $h(G_L)$ with the stable throughput region $R_T$ given in (28), and thus it is a subset of the stable throughput region $R_T$. The attempt rate $G$ should be larger than or equal to the small root $G_S$, as described in (19). The complete bounded delay region of the exponential backoff scheme is given as follows:

$$R_D = \left[ \sqrt{1-e^{-aG_S}}, h(G_L) \right). \tag{34}$$

For both access mechanisms, the bounded delay region $R_D$ of exponential backoff, as shown in Fig. 4, is a subset of the stable throughput region $R_T$. Outside this bounded delay region, i.e., $R_T \setminus R_D$, the system may still have a stable throughput but the mean delay quickly escalates to an unacceptably high level.

Under the RTS/CTS access mechanism, nodes detect the collision earlier than under the Basic access mechanism. Presumably, the channel utilization of the RTS/CTS scheme should outperform that of the Basic scheme. Fig. 4 shows that the RTS/CTS scheme has a much larger bounded delay region. Also, both Fig. 3 and Fig. 4 show that the network throughput of the RTS/CTS scheme is less sensitive to the retransmission factor $q$. In a nutshell, the RTS/CTS access mechanism offers a higher achievable throughput than the Basic access mechanism, which is consistent with the results reported in [1] and [5].

The simulation result with fixed input rate $\hat{\lambda} = 0.3$ shown in Fig. 6 verifies the analytical packet queuing delay given by (31). This figure also shows that the packet queuing delay remains small within the bounded delay region, but it increases rapidly outside this region. Another example of the bounded delay region is displayed in Fig. 7, which exhibits the mean queuing delay versus the input rate $\hat{\lambda}$ for fixed retransmission factor $q = 0.2$. It can be clearly seen that the mean queuing delay of the packet becomes unacceptably large if the aggregate input rate is on the periphery of the bounded delay region. Furthermore, as expected, both figures depict that the RTS/CTS access mechanism performs better than the Basic access mechanism.

## V. CONCLUSION

In this paper, we introduce a queuing model based on the Markov state space of the IEEE 802.11 DCF with exponential backoff scheduling algorithm. We obtain the stable throughput condition and the bounded delay condition from the characteristic equation of network throughput and service time distribution of HOL packets. The stable throughput region and bounded delay region for the two kinds of access mechanisms of 802.11 DCF protocol are determined by standard queueing analysis. Our results show that the performance of the Basic access mechanism is highly dependent on the aggregate input rate and the retransmission factor, while performance for the RTS/CTS access mechanism is only marginally sensitive to these parameters. Hence, we conclude that the RTS/CTS access mechanism is more stable and robust than its counterpart.

APPENDIX I. SERVICE TIME DISTRIBUTION FOR THE 802.11 DCF WITH EXPONENTIAL BACKOFF

The first and second moments of service time, $E[X]$ and $E[X^2]$, are listed below:

$$E[X] = a(t_A + t_S) + a\alpha_0 + at_W(1 - \alpha\alpha_0) + \frac{at_C(1-p)}{p} + \frac{a[t_W(1-\alpha)p + p]}{\alpha p(p+q-1)(1-\alpha\alpha_0 p)^{-1}}. \quad (35)$$

$$E[X^2] = C(p,q) \lim_{K \to \infty} \left[ (1-p)/q^2 \right]^{K-1} + D(p,q), \quad (36)$$

where 
$$C(p,q) = \frac{2a^2[t_W(1-\alpha)+1]^2 \left[(p+q^2-1)(p+q-1-qp+qp^2) - q^3 p^2\right]}{\alpha^2 q^2 p^2 (p+q-1)(p+q^2-1)(1-\alpha\alpha_0 p)^{-1}} \quad (37)$$

and
$$\begin{aligned} D(p,q) = a^2 \{ &(t_A + t_C + 1)(t_A + t_C)\alpha\alpha_0(1-p) + (t_A + t_W)(t_A + t_W - 1)(1-\alpha_0) \\ &+ 2p^{-2}(1-\alpha\alpha_0 p)(t_C - t_C p - p)(t_S p + t_C - pt_C) + (t_A + t_W + 1)(t_A + t_W)(1-\alpha)\alpha_0 \\ &+ (t_A + t_S + 1)(t_A + t_S)\alpha\alpha_0 p + t_C(t_C + 1)(1-p)p^{-1}(1-\alpha\alpha_0 p) + t_S(t_S + 1)(1-\alpha\alpha_0 p) \\ &+ \frac{2[t_A(1-\alpha\alpha_0 p) + t_C\alpha\alpha_0(1-p) + t_W(1-\alpha\alpha_0) + \alpha_0(1-\alpha p)]}{\alpha p(p+q-1)[\alpha(p+q-1)(t_S p + t_C - pt_C) + t_W(1-\alpha)p + p]^{-1}} \\ &+ \frac{2(t_W - \alpha t_W + 1)[(t_C + 1)(1-p) - q]}{\alpha(p+q-1)^2 (1-\alpha\alpha_0 p)^{-1}} + \frac{2(t_W - \alpha t_W + 1)(t_S p + t_C - pt_C)}{\alpha p(p+q-1)(1-\alpha\alpha_0 p)^{-1}} \\ &+ \frac{2(t_W - \alpha t_W + 1)^2 q(1-\alpha\alpha_0 p)}{\alpha^2 (p+q-1)(p+q^2-1)} + \frac{t_W(t_W + 1)(1-\alpha)}{\alpha(p+q-1)(1-\alpha\alpha_0 p)^{-1}} \} + E[X]. \end{aligned} \quad (38)$$

APPENDIX II. COMPARISONS OF PROBABILISTIC BACKOFF SCHEME AND WINDOW-BASED BACKOFF SCHEME

The analytical results of the probabilistic backoff scheme and the window-based backoff scheme for IEEE 802.11 DCF networks are provided in the Table. 2.

References

[1] G. Bianchi, "Performance Analysis of the IEEE 802.11 Distributed Coordination Function," *IEEE Journal of selected areas in COMMUN.*, Vol. 18, No. 3, pp. 535–547, 2000.

[2] P. Chatzimisios, V. Vitsas, and A. C. Boucouvalas, "Throughput and delay analysis of IEEE 802.11 protocol," in *Proc. 5th IEEE Workshop Networked Appliances*, pp. 168–174, Oct. 2003.

[3] Z. Hadzi-Velkov and B. Spasenovski, "Saturation throughput: delay analysis of IEEE 802.11 DCF in fading channel," in *Proceedings of IEEE ICC2003*, May, 2003.

[4] H. Wu, Y. Peng, K. Long, S. Cheng, and J. Ma, "Performance of reliable transport protocol over IEEE 802.11 wireless LAN: analysis and enhancement," in *Proceedings of IEEE INFOCOM2002*, June 2002.

[5] C. H. Foh and M. Zukerman, "Performance analysis of the IEEE 802.11 MAC protocol," in *Proceedings of the European Wireless 2002 Conference*, pp. 184-190, Florence, Italty, Feb. 2002.

[6] M. M. Carvalho and J. J. Garcia-Luna-Aceves, "Delay analysis of IEEE 802.11 in single-hop networks," in *Proceedings of the 11th IEEE International Conference on Network Protocols,* p. 146, Nov. 2003.

[7] T. Sakurai and H. L. Vu, "MAC access delay of IEEE 802.11 DCF," *IEEE Trans. Wireless Commun.*, vol. 6, no. 5, pp. 1702–1710, May 2007.

[8] D. Xu, T. Sakurai, and H. L. Vu, "An analysis of different backoff functions for an IEEE 802.11 WLAN,"*in Proceedings of VTC2008*, 2008

[9] M. Ergen and P. Varaiya, "Throughput analysis and admission control in IEEE 802.11a," *Mobile Networks and Applications*, vol. 10, no. 5, pp. 705–706, Oct. 2005.

[10] K. Duffy, D. Malone, and D. J. Leith, "Modeling the 802.11 distributed coordination function in non-saturated conditions," *IEEE Commun. Lett.*, vol 9, pp. 715, 2005.

[11] D. Malone, K. Duffy, and D. J. Leith, "Modeling the 802.11 distributed coordination function in nonsaturated heterogeneous conditions," *IEEE/ACM Transactions on Networking (TON)*, v.15 n.1, pp.159-172, Feb. 2007.

[12] Y. S. Liaw, A. Dadej, and A. Jayasuriya, "Performance analysis of IEEE 802.11 DCF under limited load," in *Proc. Asia-Pacific Conf. on Commun.*, vol. 1, Perth, Western Australia, Oct 2005, pp. 759–763

[13] S.-T. Cheng, M. Wu "Performance Evaluation of Ad-Hoc WLAN by M/G/1 Queueing Model," *Proc. International Conference on Information Technology: Coding and Computing*, vol. 2, pp. 681 - 686, April 2005.

[14] P. E. Engelstad and O. N. Østerbø, "Analysis of the total delay of IEEE 802.11e EDCA and 802.11 DCF," in *Proc. IEEE ICC'2006*, pp. 552-559, Istanbul, June 2006.

[15] O. Tickoo and B. Sikdar, "Queueing Analysis and Delay Mitigation in IEEE 802.11 Random Access MAC Based Wireless Networks," in *Proc. IEEE INFOCOM Conf.*, Mar. 2004.

[16] N. Bisnik and A. Abouzeid, "Queuing network models for delay analysis of multihop wireless ad hoc networks," *Ad Hoc Networks*, vol. 7, no. 1, pp. 79–97, 2009

[17] Pui King Wong, Dongjie Yin, and Tony T. Lee, "Analysis of Non-Persistent CSMA Protocols with Exponential Backoff Scheduling," *IEEE Trans. Commun.*, vol. 59, no. 8, pp.2206-2214, Aug. 2011.

[18] Pui King Wong, Dongjie Yin, and Tony T. Lee, "Performance Analysis of Markov Modulated 1-Persistent CSMA/CA Protocols with Exponential Backoff Scheduling," *Wireless Networks*, vol. 17, no. 8, pp.1763-1774, Nov. 2011.

[19] J. Goodman, A. G. Greenberg, N. Madras and P. March, "Stability of binary exponential backoff," *J. ACM*, vol. 35, pp. 579-602, 1988.

[20] J. Hastad, T. Leighton and B. Rogoff, "Analysis of backoff protocols for multiple access channels," *SIAM J. Comput.*, vol. 25, pp. 740-744, 1996.

[21] H. AL-Ammal, L. A. Goldberg and P. MacKenzie, "An improved stability bound for binary exponential backoff," *Theory Comput. Syst.*, vol. 30, pp. 229-244, 2001.

[22] H. Takagi, Queueing Analysis, *A Foundation of Performance Evaluation, Volume 3: Discrete-Time Systems*, North-holland, 1993.

[23] B-J Kwok, N-O Song, and L. E. Miller, "Analysis of the stability and performance of exponential backoff,*" IEEE WCNC*, vol.3, pp. 1754 – 1759, 2003.

[24] N.O. Song, B. J. Kwak, and L. E. Miller, "On the Stability of Exponential Backoff," *Journal of Research of the National Institute of Standards and Technology*, vol. 108, pp. 289-297, 2003.

[25] L. Kleinrock and F. A. Tobagi, "Packet Switching in Radio Channels: Part 1-Carrier Sense Multiple-Access Modes and Their Throughput-Delay Characteristics," *IEEE Trans. Commun.*, vol. COM-23, pp. 1400-1416, 1975.

[26] E. Cinlar and R. A. Agnew, "On the Superposition of Point Processes", *Journal of the Royal Statistical Society,* Series B, vol. 30, No. 3, pp. 576-581, 1968.

[27] W. Feller, Introduction to probability theory and its applications, John Wiley, New York, London, Sydney, vol. 1, 1957.

[28] Y. Yang and T. S. P. Yum, "Delay Distributions of Slotted ALOHA and CSMA," *IEEE Trans. Commun.*, vol. 51, No. 11, pp.1846-1875, 2003.

List of Figures and Tables

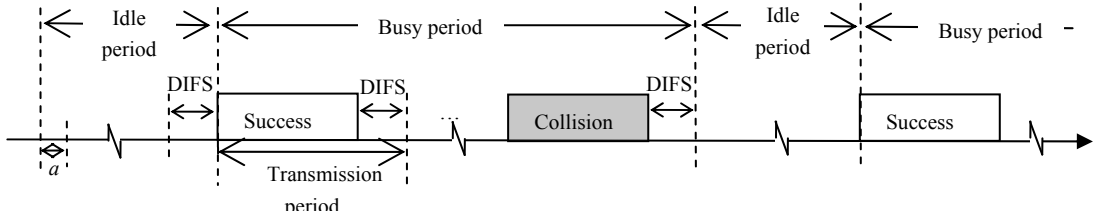

Fig. 1. The alternating renewal process of the 802.11 DCF

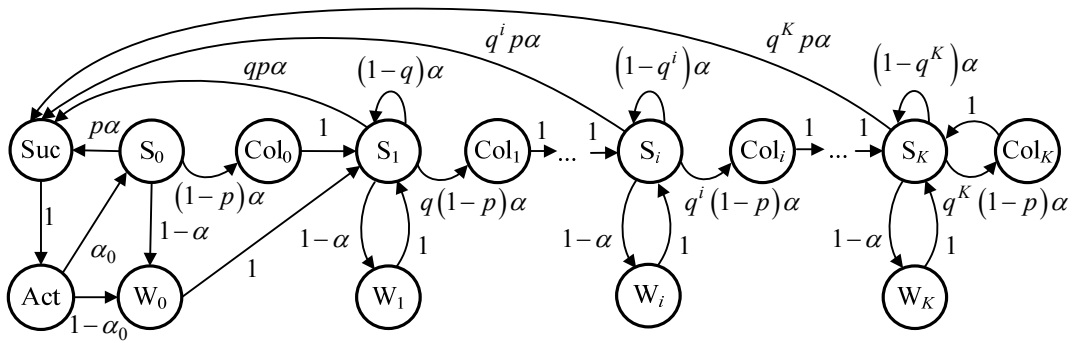

Fig. 2. Markov chain of HOL packet for IEEE 802.11 with DCF

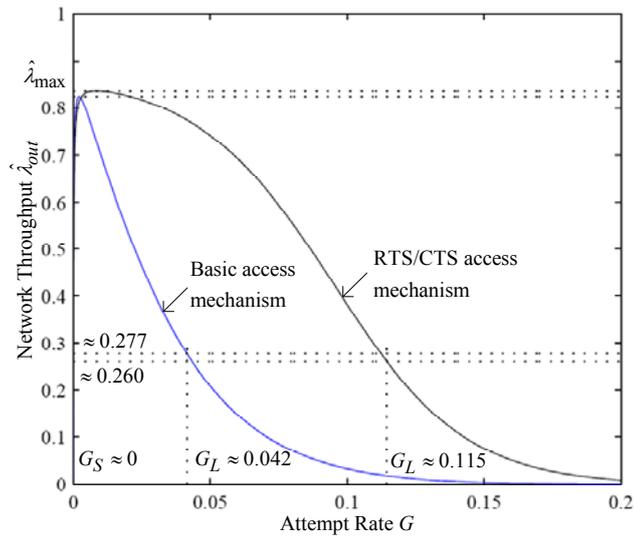

Fig. 3. Network throughput versus attempt rate $G$ of the 802.11 DCF

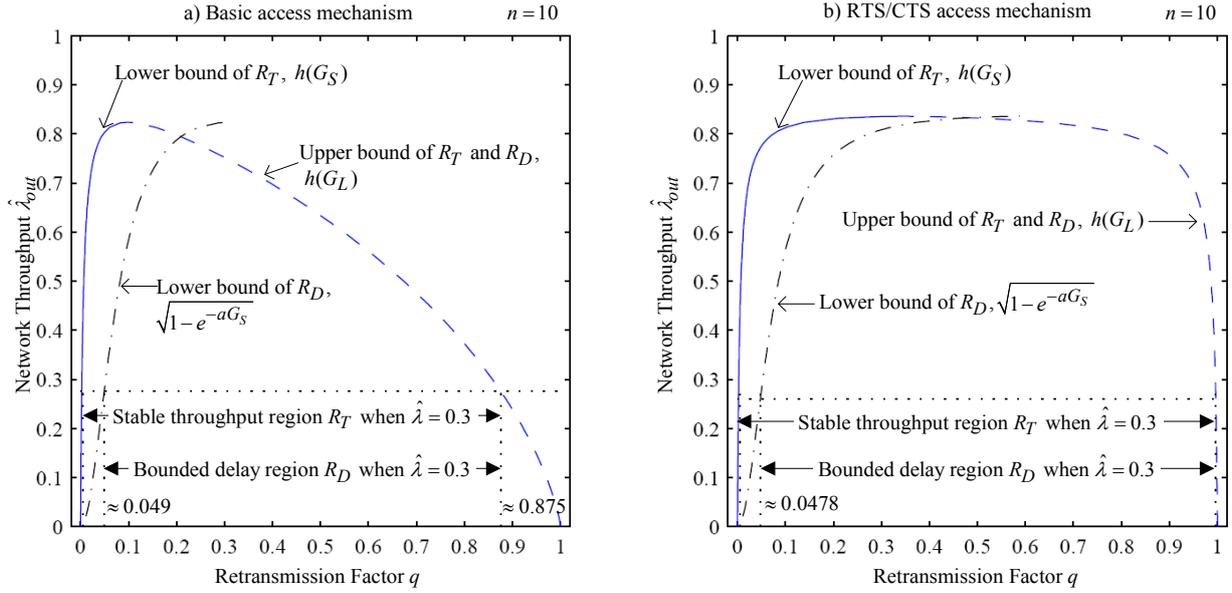

Fig. 4. Stable regions of the 802.11 DCF with exponential backoff:

a) Basic access mechanism; b) RTS/CTS access mechanism.

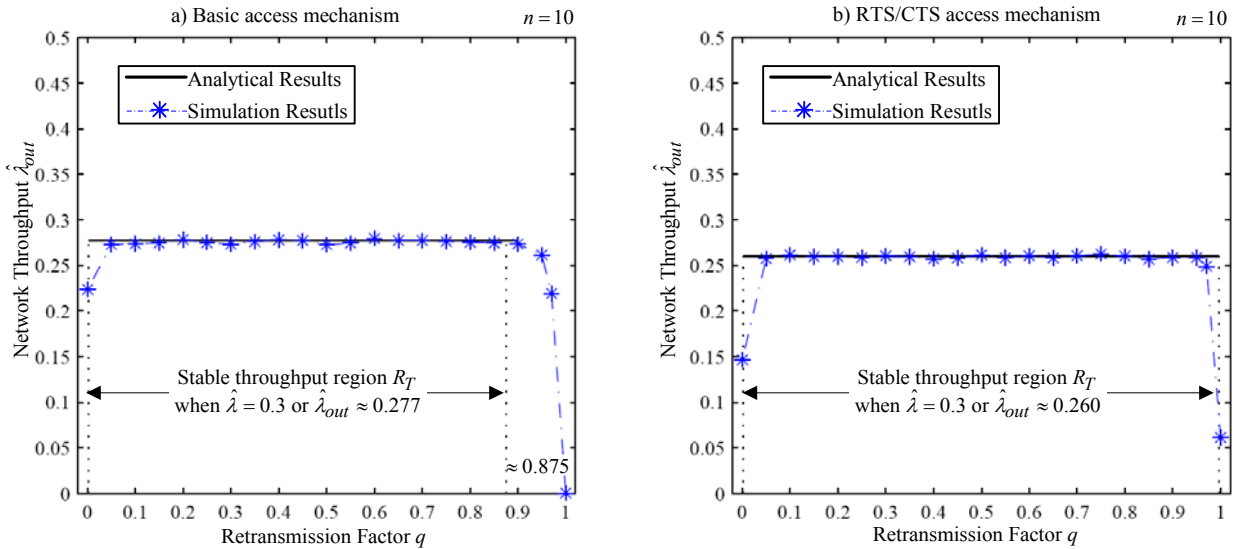

Fig. 5. Stable throughput region of exponential backoff:

a) Basic access mechanism; b) RTS/CTS access mechanism.

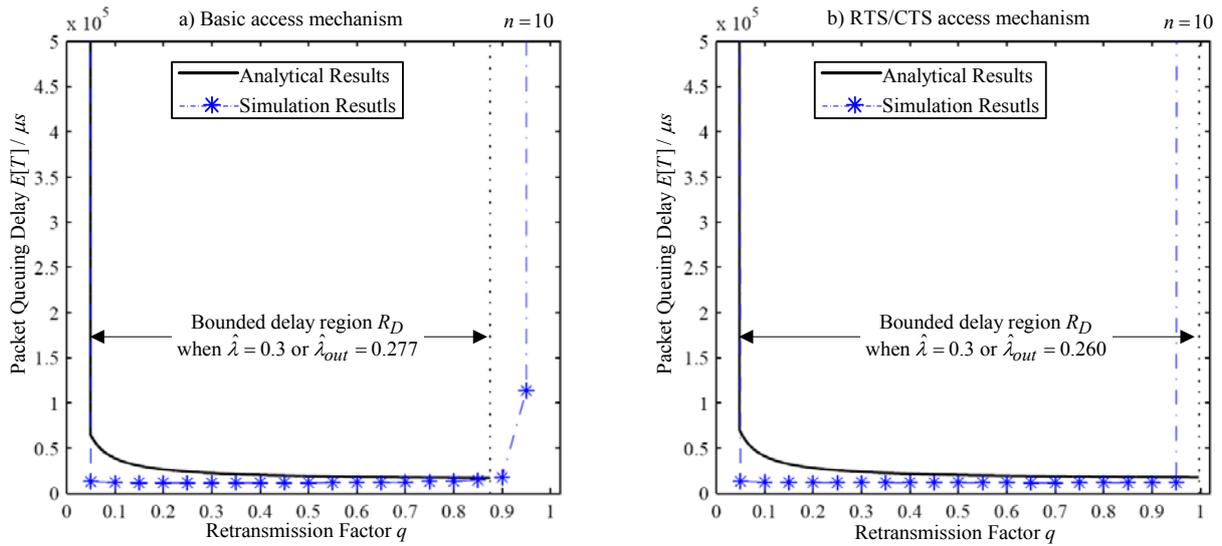

Fig. 6. Packet queuing delay versus retransmission factor of exponential backoff:

a) Basic access mechanism; b) RTS/CTS access mechanism.

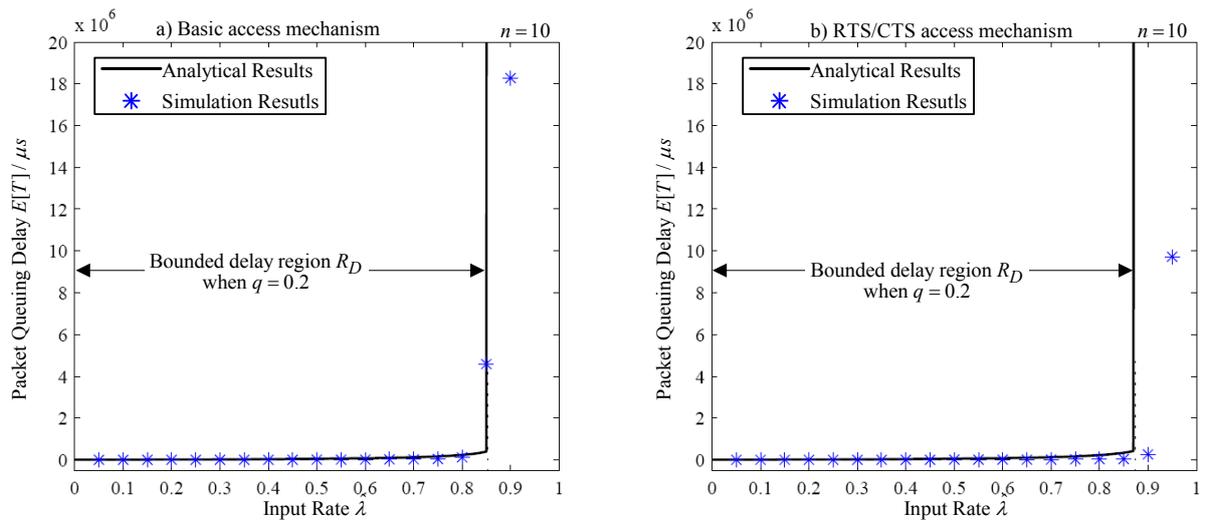

Fig. 7. Packet queuing delay versus input rate of exponential backoff:

a) Basic access mechanism; b) RTS/CTS access mechanism.

| Parameters | Time/ μs | Slot time units/ ($a$=50μs and round to integer ceiling) |
|---|---|---|
| $E[P]$ | 8184 | 164 |
| $T_D$ | 128 | 3 |
| $T_S^{bas}$ | 8982 | 180 |
| $T_C^{bas}$ | 8713 | 175 |
| $T_S^{rts}$ | 9568 | 192 |
| $T_C^{rts}$ | 417 | 9 |

Table. 1.    Values of System Parameters

| | Probabilistic backoff | Window-based backoff |
|---|---|---|
| backoff parameter | attempting with probability of $q^i$ at phase $i$ | chosen uniformly from the interval [0, $W_i$-1] at phase $i$ <br> $W_i$: contention window size at phase $i$ |
| mean holding time | $1/q^i$ at phase $i$ | $\frac{W_i+1}{2}$ at phase $i$ <br> as in [1][24][28] |
| throughput equation | $\hat{\lambda}_{out} = \frac{P_t P_S E[P]}{(1-P_t)a + P_t P_S T_S + P_t(1-P_S)T_C}$ | $S = \frac{P_t P_S E[P]}{(1-P_t)\sigma + P_t P_S T_S + P_t(1-P_S)T_C}$ <br> as in [1][10][11] <br> $S$: throughput, $\sigma$: the slot time size |
| queueing delay | exact expressions given in Appendix I | approximate expressions given in [13]–[16] |
| throughput stability | **STC:** $G_S(\hat{\lambda}_{out}) \leq G \leq G_L(\hat{\lambda}_{out})$ <br> Region: $R_T = [h(G_S), h(G_L)]$ | N/A |
| bounded queueing delay | **BDC:** $E[X^2] < \infty$ <br> Region: $R_D = \left[\sqrt{1-e^{-aG_S}}, h(G_L)\right)$ | N/A |

Table. 2.    Comparisons of the probablitistic scheme and window-based scheme